\documentstyle[12pt]{article}
\begin{document}
\begin{titlepage}
\title{The Harari-Shupe preon model\\ and\\ nonrelativistic quantum phase space}
\author{
{P. \.Zenczykowski }\\
{\em Division of Theoretical Physics},
{\em Institute of Nuclear Physics}\\
{\em Polish Academy of Sciences}\\
{\em Radzikowskiego 152,
31-342 Krak\'ow, Poland}\\
E-mail: piotr.zenczykowski@.ifj.edu.pl
}
\maketitle
\begin{abstract}
We propose that the whole algebraic structure
 of the Harari-Shupe rishon model originates
 via a Dirac-like linearization of quadratic form 
${\bf x}^2+{\bf p}^2$, with position and momentum satisfying standard commutation relations.
The scheme does not invoke the concept of preons as spin-1/2
subparticles, thus evading the problem of preon confinement, 
while fully explaining all symmetries emboded in the Harari-Shupe model.
Furthermore, the concept of quark colour is naturally linked to the
ordering of rishons.
Our scheme leads to group $U(1) \otimes SU(3)$ combined with $SU(2)$, with
two of the SU(2) generators not commuting with reflections.
An interpretation of intra-generation quark-lepton transformations in terms of genuine 
rotations and reflections in phase space is proposed.
\end{abstract}
\vfill
\end{titlepage}

\section{Introduction}

The Standard Model (SM) is  very successful  in its 
description
of the
interactions of elementary particles. Yet, putting its successes aside,
it contains many seemingly arbitrary features which indicate the need
for a deeper explanation.
In particular, while it was very natural to assume that the 
gauge principle known from electromagnetism should be extended 
to other interactions,
the choice of $U(1)\otimes SU(3) \otimes
SU(2)_L$  as the gauge group is dictated solely by experiment and 
remains unexplained at the level of theoretical principles.
In other words, we do not know a simple theoretical reason that presumably
underlies the emergence of internal symmetries and could explain
the structure of SM
generation. \\
   
Following the success of composite models throughout the history of physics,
 the
proliferation of fundamental particles naturally
led people to consider quarks and leptons
as built of some constituents, dubbed ``preons'' 
by Pati and Salam \cite{PatiSalam1974}. 
The most interesting of such models is the
Harari-Shupe model \cite{Harari}, which describes the structure of a single SM
generation with the help of only two spin-$1/2$ ``rishons'' $V$ and $T$,
 of charges
0 and +1/3 respectively. This is shown in Table \ref{table1}, where total 
charges and hypercharges of particles are also listed
(for other preon models, see e.g. \cite{Fritzsch}.)
However, though algebraically very economical, the rishon model
has several drawbacks.
These include: the issue of preon confinement at extremely
small distance scales (when confronted
with the uncertainty principle), the apparent
absence of spin 3/2 fundamental particles, and the lack of explanation 
as to why the
ordering of three rishons is important 
(this ordering gives rise to the ``threeness'' of the colour degree of freedom).
These problems were addressed e.g. in \cite{HarariSeiberg}.\\

\begin{table}[b]
\caption{Rishon structure of leptons and quarks with a third component of weak
isospin $I_3=+1/2$}
\label{table1}
\begin{center}
\begin{math}
\begin{array}{c|cccc|cccc}
&~~~\nu_e~~~&~~u_R~~&~~u_G~~&~~u_B~~&~~e^+~~&
~~\bar{d}_R~~&~~\bar{d}_G~~&~~\bar{d}_B~~\\
\hline 
&VVV&VTT&TVT&TTV&TTT&TVV&VTV&VVT\rule{0mm}{5mm}\\
Q &0&+\frac{2}{3}&+\frac{2}{3}&+\frac{2}{3}&+1
&+\frac{1}{3}&+\frac{1}{3}&+\frac{1}{3}
\rule{0mm}{5mm}\\
Y &-1&+\frac{1}{3}&+\frac{1}{3}&+\frac{1}{3}&+1
&-\frac{1}{3}&-\frac{1}{3}&-\frac{1}{3}
\rule{0mm}{5mm}

\end{array}
\end{math}
\end{center}
\end{table}

On the other hand, one has to be aware that 
explaining the existence of a multiplet of some symmetry 
in terms of particle constituents,
though so successful in the old days, may be going too far. For example,
Heisenberg was against such unwarranted ``explanations'' 
of symmetries \cite{Heisenberg}.
His point of view was that at some point in the process of dividing
matter again and again, the very concept of ``dividing'' loses its meaning.\\
 
Recently, a proposal along such general lines has been made by Bilson-Thompson, who
suggests correspondence between the algebraic structure of the Harari-Shupe
model and the topological properties of braids composed of three ``helons''
\cite{BT}. 
In his
model, the ``binding''
of preons is topological in nature, and thus preons are not to be considered as
confined point-like particles.\\

The approach presented below belongs to this very general line of reasoning with
Harari-Shupe rishons considered to be purely algebraic ``components'',  
and not ordinary confined point-like particles
(the meaning of the term ``algebraic component'' will be fully explained 
as we proceed with the presentation of our proposal).
      
\section{Spatial and internal symmetries}
When searching for a principle underlying the appearance of quantum numbers
corresponding to internal symmetries, one should note that some quantum
attributes of elementary particles are clearly associated with 
the properties of
classical macroscopic continuous space in which these particles move 
(e.g. spin). This suggests that internal quantum numbers could perhaps 
be also connected with the properties of some properly understood 
``classical space''. 
Such a
point of view is held by several physicists, e.g. Penrose, 
who writes in \cite{Penrose}: ``I do not believe that a real understanding of the
nature of elementary particles can ever be achieved without a simultaneous
deeper understanding of the nature of spacetime itself''.\\

Now, it should be noted that all the quantum numbers for which a connection
with macroscopic classical space is known can be established using strictly
nonrelativistic reasoning.
This refers not only to spin and parity, but also to the existence of particles
and antiparticles (and thus C-parity).
Indeed, with antiparticles interpreted as particles moving backwards in time, it
should be obvious that
the existence of these two groups of objects 
is closely related to the existence of the operation of time reflection, and
not to truly relativistic transformations.
It may be formally shown that antiparticles also emerge  
when the strictly nonrelativistic
Schr\"odinger equation is linearised \'a la Dirac \cite{HK2003}.\\

It should be also kept in mind that
 the connection between space and time is more subtle than 
   the standard mathematical form of special relativity would
suggest. 
Indeed, the latter form emerges only when the Einstein radiolocation prescription 
for the synchronization of distant clocks is adopted. 
However, distant clocks may be
synchronised in various ways, reflecting the presence of a kind of gauge
freedom related to the impossibility of measuring the one-way speed of light.
With a suitable gauge even absolute simultaneity may be achieved, obviously
without spoiling the successes of the theory of special relativity.
\cite{Poincare}.\\

  In view of the nonrelativistic origin of all quantum
numbers for which their connection with the macroscopic classical arena has been
established, {\it the simplest expectation is that other observed quantum numbers of
elementary particles may be also inferred through nonrelativistic
reasoning} \cite{APPB2007a}.

\section{Phase space as arena of physical processes}
After restricting our considerations to the nonrelativistic approach, 
we observe that  
the description of the time evolution of a single particle
may be provided
 either on the background of
   the three-dimensional position space, 
   or on that of the six-dimensional phase space.
   Indeed, the Hamiltonian formalism - where 
   position and momentum are {\it independent} variables - suggests
 that we may
   treat phase space as an arena of physical events.
   We now recall that our goal is to understand the origin
   of the quantum numbers of elementary particles, and that
   quantum mechanics works in phase space. Consequently, 
   the choice of phase space 
   as an arena for physical processes  
   seems to be a proper choice for our purposes.\\
   
Consideration of nonrelativistic phase space as the arena of events permits 
a generalization of ordinary transformations of space to 
those of phase space.
Obviously, if such generalized  transformations 
are to be feasible, one has to add
another physical constant, of dimension [momentum/position], 
which permits the
expression of all six independent phase-space 
coordinates in terms of the same dimensional units.
The actual value of this constant is completely irrelevant at this moment.
It suffices to say here
that - together with the Planck constant (and the velocity of light $c$) - a natural mass scale
is then set. The
introduction of such a constant was considered by many, 
in particular by Born,
who observed that various physical quantities are invariant under 
the so-called
``reciprocity'' transformations  
${\bf x} \to {\bf p}$, ${\bf p} \to -{\bf x}$ \cite{Born1949}.\\

The choice of phase space as an arena is possible because physics does not deal 
with reality ``directly'',
providing only its descriptions instead.
Consequently, 
different descriptions may be used to deal with the same physics, leading
to the same (or similar) predictions.
A well-known example of this general truth is provided by gauge theories, 
whose physical predictions are 
independent of the gauge.

\section{Basic invariant and its linearization}
The basic invariant in the standard description 
(3D arena of positions) 
is ${\bf x}^2$.
In the phase-space-based description we have to consider ${\bf p}^2$ as well, which
constitutes another fully independent invariant of this kind. 
If we want to maintain maximal
symmetry between position and momentum, 
then only the combination ${\bf x}^2+{\bf p}^2$ is admitted 
as a possible invariant in phase
space (as considered also by Born \cite{Born1949}), with the relevant invariance
group being $O(6)$. 
\\

We now consider ${\bf x}$ and ${\bf p}$ to be operators satisfying
standard position-momentum commutation relations. When one requires 
restriction to the subgroup of $O(6)$ under which these commutation
relations stay invariant, the resulting symmetry group is $U(1) \otimes SU(3)$,
as is well known from the case of the standard 3D 
harmonic oscillator. The $U(1)$ factor takes care of the Born reciprocity
transformations (${\bf x'}=-{\bf p},~{\bf p'}=+{\bf x}$) 
and their squares, i.e. ordinary reflections 
(${\bf x''}=-{\bf p'}=-{\bf x},~{\bf p''}=+{\bf x'}=-{\bf p}$), while $SU(3)$
constitutes a generalization of the ordinary rotation group $SO(3)$ 
\cite{APPB2007a}.
The generator of $U(1)$ in phase space is
\begin{equation}
R^z={\bf x}^2+{\bf p}^2,
\end{equation}
where superscript $z$ collectively denotes $({\bf p},{\bf x})$ and indicates 
that we are dealing with the representation in
phase space.\\

Let us now introduce the crucial step of our approach, 
i.e. the linearization of ${\bf x}^2+{\bf p}^2$
\'a la Dirac.
We achieve this by considering the square of
\begin{equation}
\label{Ap+Bx}
{\bf A}\cdot {\bf p}+{\bf B}\cdot {\bf x},
\end{equation}
with matrices ${\bf A}$, ${\bf B}$ satisfying standard anticommutation relations:
\begin{eqnarray}
\{A_k,A_l\}=\{B_k,B_l\}=2\delta_{kl},&~~~~~~~~~~&\{A_k,B_l\}=0.
\end{eqnarray}
We shall use the following representation:
\begin{eqnarray}
A_k=\sigma_k \otimes \sigma_0 \otimes \sigma_1 &~~~~~~~~
&B_k=\sigma_0 \otimes \sigma_k \otimes \sigma_2.
\end{eqnarray}
Then the seventh anticommuting matrix of the Clifford algebra generated by
${\bf A}$ and ${\bf B}$ is 
\begin{equation}
B=iA_1A_2A_3B_1B_2B_3=\sigma_0 \otimes \sigma_0 \otimes \sigma_3. 
\end{equation}
One finds:
\begin{equation}
({\bf A}\cdot {\bf p}+{\bf B}\cdot {\bf x})
({\bf A}\cdot {\bf p}+{\bf B}\cdot {\bf x})=
R^z+R^{\sigma}\equiv R,
\end{equation}
with (superscript $\sigma$ refers to matrix space)
\begin{equation}
R^{\sigma}=\sum_k \sigma_k \otimes \sigma_k \otimes \sigma_3 \equiv \sum_k
R^{\sigma}_k
\end{equation}
appearing here because ${\bf x}$ and ${\bf p}$ do not commute.
Thus, operator $R$ constitutes the total $U(1)$ generator, 
a sum of contributions $R^z$
from phase space and $R^{\sigma}$ from matrix space.
To proceed further, we find the eigenvalues of $R^z$ and $R^{\sigma}$.
The eigenvalues of $R^z$ are obviously $3,5,7,..$, 
while for $R^{\sigma}$ there are eight eigenvalues:
$-3,+1,+1,+1,-1,-1,-1,+3$.
The lowest eigenvalue of $R^z$ is $+3$, and
we shall adopt it in the following as
corresponding to some ``vacuum''.
The lowest absolute value of $R^{\sigma}$ is $+1$, i.e. it is $smaller$ than the
minimal value
of $+3$
allowed in the standard 3D harmonic oscillator.
We shall discuss the meaning of this low value further on.

\section{Recovering the Harari-Shupe model}
We now adopt the lowest ``no-excitation'' value of $+3$ for $R^z$, and propose to identify
\begin{equation}
\label{GMNGa}
Q=\frac{1}{6}R=+\frac{1}{2}+\frac{1}{6}R^{\sigma}
\end{equation}
with the charge operator for the set of 
$\nu_e,u_R,u_G, u_B, e^+, \bar{d}_R,\bar{d}_G,\bar{d}_B$
shown in Table \ref{table1}. The
second term on the r.h.s. in Eq. (\ref{GMNGa})  obviously corresponds to the
hypercharge $Y$ in the Gell-Mann-Nishijima-Glashow formula \cite{Glashow} 
$Q=I_3+Y/2$ (with $I_3=+1/2$) if we
identify:
\begin{equation}
Y=\frac{1}{3}R^{\sigma}.
\end{equation}
In order to see strict correspondence with the rishon model, we introduce
\begin{equation}
Y_k=\frac{1}{3}R^{\sigma}_k=\frac{1}{3}\sigma_k \otimes \sigma_k \otimes \sigma_3.
\end{equation}
Since all commutators of $Y_k$ with themselves vanish:
\begin{equation}
[Y_k,Y_l]=0,
\end{equation}
it follows that $Y_1,Y_2,Y_3$ and $Y$ may be simultaneously diagonalized.
The eigenvalues of $Y_k$ are $\pm 1/3$. Thus, we have
$2^3=8$ possibilities for $Y=Y_1+Y_2+Y_3$, as shown in Table \ref{table2}.
Strict correspondence with the rishon model is obvious.
The value $Y_k=-1/3~ (+1/3)$ corresponds to rishon $V$ ($T$), 
while the position of the rishon  
corresponds to the value of $k$.
Thus $VTT$, corresponding to $(Y_1,Y_2,Y_3)=(-1/3,+1/3,+1/3)$ is clearly
different from $TVT$ corresponding to   
 $(Y_1,Y_2,Y_3)=(+1/3,-1/3,+1/3)$, etc.
 In addition, there is no need for any ``dynamical'' preon confinement, as in our
 scheme 
    the structure identified by Harari and Shupe corresponds 
to a mere group-theoretical procedure of 
adding the three components of Y.
 The antiparticles of 
$\nu_e,u, e^+,\bar{d}$, 
i.e. $\bar{\nu}_e,\bar{u}, e^-,d$ (all of them with $I_3=-1/2$) are
described by the complex-conjugate representation.
One then finds \cite{APPB2007a} that
$Q = -1/2+R^{\sigma}/6$
(the sets of eigenvalues of $R^{\sigma}$ and $-R^{\sigma}$ being identical), 
with $Y,Y_k$ effectively changing their signs, 
exactly as in the Harari-Shupe model.\\

\begin{table}
\caption{Decomposition of eigenvalue of Y into eigenvalues of its
components}
\label{table2}
\begin{center}
\begin{math}
\begin{array}{c|cccc|cccc}
&~~~\nu_e~~~&~~u_R~~&~~u_G~~&~~u_B~~&~~e^+~~&
~~\bar{d}_R~~&~~\bar{d}_G~~&~~\bar{d}_B~~\\
\hline 
Y &-1&+\frac{1}{3}&+\frac{1}{3}&+\frac{1}{3}&+1
&-\frac{1}{3}&-\frac{1}{3}&-\frac{1}{3}
\rule{0mm}{5mm}\\
\hline 
Y_1 &-\frac{1}{3}&-\frac{1}{3}&+\frac{1}{3}&+\frac{1}{3}&+\frac{1}{3}
&+\frac{1}{3}&-\frac{1}{3}&-\frac{1}{3}
\rule{0mm}{5mm}\\
Y_2 &-\frac{1}{3}&+\frac{1}{3}&-\frac{1}{3}&+\frac{1}{3}&+\frac{1}{3}
&-\frac{1}{3}&+\frac{1}{3}&-\frac{1}{3}
\rule{0mm}{5mm}\\
Y_3 &-\frac{1}{3}&+\frac{1}{3}&+\frac{1}{3}&-\frac{1}{3}&+\frac{1}{3}
&-\frac{1}{3}&-\frac{1}{3}&+\frac{1}{3}
\rule{0mm}{5mm}\\
\hline
{\rm colour} \# & ~0 & ~1 & ~2 & ~3 & ~0
&~1 &~2 &~3
\rule{0mm}{5mm}
\end{array}
\end{math}
\end{center}
\end{table}

The labelling of the algebraic components gives rise to colour and $SU(3)$. Indeed,
the nine generators of $U(1)\otimes SU(3)$ are represented
in our Clifford algebra by the $U(1)$ generator $R^{\sigma}$ 
and the $SU(3)$ generators $F^{\sigma}_a$ ($a=1,2.,,,8$),  
formed in the standard way as eight appropriate combinations of
antisymmetric products of $A_k$ and $B_l$,
whose explicit form
is given in \cite{APPB2007a}. 
Using this explicit form, it is straightforward
to calculate that
\begin{equation}
\sum_{a=1}^8(F^{\sigma}_a)^2=4 ~(1+\frac{1}{3}\sum_k \sigma_k\otimes
\sigma_k\otimes \sigma_0)=4(1+YB)
\end{equation}\\
Since $[Y,B]=[Y,YB]=[B,YB]=0$, the matrix $YB$ may be diagonalized
simultaneously with $Y$ and $B$. The eigenvalues of $YB$ are $-1,+1/3,+1/3,+1/3$,
corresponding to the $Y$ eigenvalues of $-1,+1/3,+1/3,+1/3$ (for $B=+1$)
and $Y$ eigenvalues  of $+1,-1/3,-1/3,-1/3$ (for $B=-1$).
Thus, for leptons ($YB=-1$) one has $\sum_{a=1}^8(F^{\sigma}_a)^2=0$, while
for quarks ($YB=+1/3$) one has $\sum_{a=1}^8(F^{\sigma}_a)^2=16/3$.
In our normalization of $SU(3)$ generators, this means that leptons are singlets
and quarks are triplets of $SU(3)$.

\section{Weak isospin}
In order to treat isospin in a standard way, one needs to put together 
$\nu_e,u_R,u_G, u_B$ and $e^-, {d}_R,{d}_G,{d}_B$ instead of 
$\nu_e,u_R,u_G, u_B$, and $e^+, \bar{d}_R,\bar{d}_G,\bar{d}_B$.
Since the eigenvalues of $YB$ are just $-1,+1/3,+1/3,+1/3$, as needed for
both $\nu_e, u$ and $e^-, d$, it follows that  we may use 
matrix $B$ to this end.
In fact, within our Clifford algebra there are only four matrices which commute 
with the $U(1)\otimes SU(3)$ generators.
These are: the unit matrix, $Y$, $B$, and $YB$.
Thus, in our (minimal) scheme we have
\begin{equation}
I_3=\frac{1}{2}B=\frac{1}{2}\, \sigma_0 \otimes\sigma_0 \otimes \sigma_3.
\end{equation}
The $SU(2)$ counterparts of $I_3$, i.e.
$I_k=\frac{1}{2}\,\sigma_0\otimes\sigma_0\otimes\sigma_k$ ($k=1,2$), 
commute with $YB$. They do not commute with the  
generators $F^{\sigma}_a$ of the original $SU(3)$ for
$a=1,3,4,6,8$ \cite{APPB2007a}. However, one can 
modify $F^{\sigma}_a$'s by setting $\tilde{F}^{\sigma}_a=F^{\sigma}_a$ for
$a=2,5,7$ (ordinary rotations) and $\tilde{F}^{\sigma}_a=F^{\sigma}_aB$ for the 
remaining values of $a$. Then, the $\tilde{F}^{\sigma}_a$'s still satisfy 
the $SU(3)$ commutation relations, while commuting 
with all $SU(2)$ generators \cite{APPB2007a}. 
In our scheme, the reflection operator $P^{\sigma}$ is 
obtained as a particular rotation generated by $R^{\sigma}$ \cite{APPB2007a}:
\begin{equation}
P^{\sigma}=\exp (-i\frac{\pi}{2}R^{\sigma})
\end{equation}
and it turns out to be
proportional to $I_3$.
Thus, two of the $SU(2)$ generators do not commute with reflections.
While the situation in the real world is certainly much more complex, this 
lack of commutativity seems
to be an interesting byproduct of our approach.
The generators $F^{\sigma}_a$ (or $\tilde{F}^{\sigma}_a$) obviously
commute with the reflection operator $P^{\sigma}$.
In conclusion, our scheme leads to $U(1)\otimes SU(3)$ combined with $SU(2)$.
While for $\tilde{F^{\sigma}_a}$ the two groups: $SU(3)$ and ${SU(2)}$  
also form a direct product, $U(1)$ and $SU(2)$ do not.

\section{Genuine $SU(4)$ transformations}
The appearance of the eigenvalues of $R^{\sigma}$ equal to $\pm 1$, i.e.
smaller in the absolute value
than the minimal value of $+3$ allowed by the 3D harmonic oscillator, requires
explanation in the phase-space language.
We will now show that such low eigenvalues correspond to quark
position-momentum commutation relations having been modified 
when compared to those in the lepton case.
In order to see this, we need to find transformations from the lepton sector
to the quark sector. To this end,
let us consider six ``genuine'' $SU(4)$ generators
$F^{\sigma}_{\pm n}$ ($n=1,2,3$) that - together with the nine generators of
$U(1)\otimes SU(3)$ - form fifteen rotation generators in our Clifford algebra
\cite{APPB2007b}:
\begin{eqnarray}
F_{+n}^{\sigma}&=&\frac{1}{2} \,\epsilon_{nkl}\,\sigma_k \otimes \sigma _l \otimes
\sigma _3\\
F_{-n}^{\sigma}&=&\frac{1}{2}\,(\sigma_0 \otimes \sigma _n-
\sigma_n\otimes\sigma_0)\otimes \sigma_0.
\end{eqnarray}
We shall study transformations of $Y_kB\equiv \frac{1}{3}y_k\otimes \sigma_0$ and
 $YB\equiv \frac{1}{3}y\otimes \sigma_0$ induced by finite rotations generated
 by $F^{\sigma}_{\pm n}$.

Before the transformation, $y_k$'s ($y_k=\sigma_k\otimes\sigma_k$) 
diagonalize (simultaneously) as follows: 
\begin{eqnarray}
y_1
& \to & \left[ \begin{array}{cccc}
-1&&&\\
&+1&&\\
&&-1&\\
&&&+1
\end{array} \right],\nonumber\\
y_2
 & \to & \left[ \begin{array}{cccc}
+1&&&\\
&+1&&\\
&&-1&\\
&&&-1
\end{array} \right],\nonumber\\
\label{ykdiagonalized}
y_3
& 
\to & \left[ \begin{array}{cccc}
+1&&&\\
&-1&&\\
&&-1&\\
&&&+1
\end{array} \right],
\end{eqnarray}
so that
\begin{eqnarray}
\label{oldcolours}
y&\to& \left[ \begin{array}{cccc}
+1&&&\\
&+1&&\\
&&-3&\\
&&&+1
\end{array} \right],
\phantom{xxxx}
\leftarrow
{{\rm colour}~\#'s }
\phantom{x}
\left\{\begin{array}{l}
1\\
3\\
0~{\rm (lepton)}\\
2
\end{array}\right.,\phantom{xx}
\end{eqnarray}

As an example, we focus here on $F^{\sigma}_{-2}$-generated 
rotations :
\begin{equation}
\label{F2}
\tilde{Y}_k\tilde{B}=
e^{+i\phi F^{\sigma}_{-2}}Y_kB e^{-i\phi F^{\sigma}_{-2}}.
\end{equation} 
for $\phi=\pm \pi/2$
(for the general case and 
for rotations generated by $F^{\sigma}_{+n}$ see \cite{APPB2007b}). 

After the above transformation, the $\tilde{y}_k$'s diagonalize as:
\begin{eqnarray}
\tilde{y}_1 & \to & \left[ \begin{array}{cccc}
-1&&&\\
&+1&&\\
&&+1&\\
&&&-1
\end{array} \right],\nonumber\\
\tilde{y}_2 & \to & \left[ \begin{array}{cccc}
+1&&&\\
&+1&&\\
&&-1&\\
&&&-1
\end{array} \right],\nonumber\\
\label{yprimekdiagonalized}
\tilde{y}_3& \to & \left[ \begin{array}{cccc}
+1&&&\\
&-1&&\\
&&+1&\\
&&&-1
\end{array} \right],
\end{eqnarray}
and
\begin{eqnarray}
\label{newcolours}
\tilde{y}&\to& \left[ \begin{array}{cccc}
+1&&&\\
&+1&&\\
&&+1&\\
&&&-3
\end{array} \right],
\phantom{xxxx}
\leftarrow
{{\rm colour}~\#'s }
\phantom{x}
\left\{\begin{array}{l}
1\\
3\\
2\\
0~{\rm (lepton)}
\end{array}\right..\phantom{xx}
\end{eqnarray}
Thus, for $\phi=\pm \pi/2$, transformation (\ref{F2}) interchanges the lepton 
with the quark of colour \# 2,
while leaving the remaining two quark colours unchanged.
Rotations by $\pm\pi/2$ generated by$F^{\sigma}_{+2}$ lead to the same result 
(see \cite{APPB2007b}).

\section{Phase-space interpretation of colour}
The meaning of the quark-lepton interchange of the previous section
may be understood in terms of
phase-space concepts through analyzing the invariance of 
expression
${\bf A}\cdot {\bf p}+{\bf B}\cdot {\bf x}$ under $F^{\sigma,z}_{\pm
n}$-generated transformations.
The $F^{\sigma,z}_{-n}$-generated transformation corresponds to a rotation
in the position space {\em relative} to the momentum space.
Thus, if one chooses to work in momentum representation of the standard 3D picture, in which
the ${\bf B}\cdot {\bf x}$ term is not present, the ${\bf A}\cdot {\bf p}$
term does not change when going from the lepton sector to the quark sector.
Hence, the same connection between the (algebraic) spin and momentum should
exist for both lepton and
quarks.
However, the connection between position and momentum gets modified. In fact,
the phase-space counterpart of Eq. (\ref{F2}) leads for general $\phi$ to new
momenta $\tilde{\bf p}$ and positions $\tilde{\bf x}$ satisfying the following
commutation relations \cite{APPB2007b}:
\begin{eqnarray}
[\tilde{x}_k,\tilde{x}_l]=[\tilde{p}_k,\tilde{p}_l]&=&0\\
\label{tildexp}
{}[\tilde{x}_k,\tilde{p}_l]~~~~~~&=&i\Delta_{kl}
\end{eqnarray}
with
\begin{equation}
\Delta=\left[
\begin{array}{ccc}
\cos 2\phi&0&\sin 2\phi\\
0&1&0\\
-\sin 2 \phi &0&\cos 2 \phi
\end{array}
\right].
\end{equation}
Commutation relations (\ref{tildexp}) become {\it diagonal} 
if 
\begin{equation}
\phi = 0, \pm \pi/2, \pm \pi, \pm 3\pi/2,.... 
\end{equation}
The cases with $\phi=0,\pm \pi$ are trivial (the latter being equivalent to
ordinary rotation by $\pm \pi$ around the second axis), 
and since $3\pi/2=\pi/2+\pi$, only
 $\phi=\pm \pi/2$ is of real interest.
This is the case of the quark-lepton interchange from Eq. (\ref{oldcolours}) 
to Eq. (\ref{newcolours}).
A similar conclusion is reached when the $F^{\sigma,z}_{+2}$-generated rotations
 are considered.

In our scheme, therefore, transformations between a lepton and three quarks
(with the same $I_3$) 
correspond to transformations
between four forms of position-momentum commutation relations:
\begin{equation}
[x_k,p_l]=i \Delta _{kl}
\end{equation}
with four different possibilities for {\it diagonal} $\Delta_{kl}$:
 \begin{equation}
 \label{gendiag2}
 \left[
\begin{array}{ccc}
+1 &0&0\\
0&+1&0\\
0 &0&+1
\end{array}
\right],\phantom{x}
\left[
\begin{array}{ccc}
+1 &0&0\\
0&-1&0\\
0 &0&-1
\end{array}
\right],\phantom{x}
\left[
\begin{array}{ccc}
-1 &0&0\\
0&+1&0\\
0 &0&-1
\end{array}
\right],\phantom{x}
\left[
\begin{array}{ccc}
-1 &0&0\\
0&-1&0\\
0 &0&+1
\end{array}
\right]\phantom{xxx}
\end{equation}
and the standard meaning of positions and momenta. Since going from
a lepton to any of the three types of quarks requires 
a (particular) genuine rotation 
in phase space, this transformation
cannot be effected in our ordinary 3D world.
\\

The fact that none of the three additional sets of commutation relations 
above is  
rotationally invariant does not entitle us to dismiss 
the presented approach, as
the argument below indicates.
The point is that in the real world we 
{\it never probe individual quarks}. Instead, 
we always probe quark aggregates, i.e. hadrons.

This is reflected also in the description provided by the Standard Model, 
in which photons or weak bosons couple to $SU(3)$-{\it singlet} quark currents, i.e.
to $\bar{q}...q$ bilinears {\it summed} over colour, or, in other words, to
objects with meson-like (and not quark-like) properties.
Thus, the SM description does not allow us to ``see'' a quark of a fixed colour.

We expect this general qualitative idea to work in our case as well. 
Our scheme certainly admits the
formation of $SU(3)$-singlets (and $SO(3)$ scalars) out of $SU(3)$-triplets.   
It has to be studied
further whether in our description, which is richer than the standard
3D one, quark aggregates of the expected properties can be 
constructed (presumably in the form of appropriate combinations of 
tensor products). 
In other words, the question is
whether one can make our quarks 
``conspire'' in such a way that the resulting aggregate
 - as a whole - behaves in a proper way under rotations.
Such a study obviously touches on the issue of confinement and
is beyond the scope of the present paper.
However, since the scheme includes the rotation group
(and, consequently, must involve all its representations), 
a positive answer seems quite
possible here. Furthermore,
it has to be stressed
 that the conceptual basis of the approach 
- i.e. the choice of phase space as the arena of physical processes, combined
with the
introduction of more symmetry between momentum and position,
 and linearization \'a la Dirac - looks so natural that it certainly
 justifies further
 studies.

\section{Reflections in phase space - isospin}
Transition between sectors of different $I_3=\pm 1/2$
is achieved by
\begin{equation}
\label{reflection1}
\tilde{X}=I_{\pm} X I^{-1}_{\pm},
\end{equation}
with $I_{\pm}$ satisfying $I_3=-I_{\pm}I_3I^{-1}_{\pm}$. 
We may take $I_{\pm}=\sigma_0\otimes\sigma_0\otimes \sigma_1$, whence
$\tilde{{\bf A}}={\bf A}$ and $\tilde{{\bf B}}=-{\bf B}$. The
invariance of 
expression
${\bf A}\cdot {\bf p}+{\bf B}\cdot {\bf x}$ requires then that the corresponding
transformation in phase space be:
\begin{eqnarray}
\tilde{{\bf p}}={\bf p}&~~~~~&\tilde{{\bf x}}=-{\bf x},
\end{eqnarray}
i.e. we get reflection in six dimensions.
Under the transformation of Eq. (\ref{reflection1}) the sign of 
the imaginary number $i$ is unchanged.
Consequently, the original commutation relations
\begin{equation}
\label{plusi}
[x_k,p_l]=+i\Delta_{kl}
\end{equation}
are replaced with
\begin{equation}
\label{minusi}
[x_k,p_l]=-i\Delta_{kl}
\end{equation}
The operation leading from Eq (\ref{plusi}) to Eq (\ref{minusi}) is not the same as complex conjugation since the latter changes 
the sign of both $p_k=-i\frac{d}{dx_k}$ and $i$, while leaving (real) $x_l$ untouched.
The two possibilities of Eqs (\ref{plusi},\ref{minusi}) exist
because the imaginary unit which is to appear on the r.h.s of
position-momentum commutation relations may be arbitrarily chosen as $+i$ or
$-i$.
With $\Delta\to -\Delta$ the four cases of Eq. (\ref{gendiag2}) are now extended to eight,
 thus exhausting all possibilities.

\section{Conclusions}
We have proposed that the Harari-Shupe  
model should be understood solely in
terms of the built-in symmetry, 
without the need to introduce ``confined preons''.
This symmetry has been shown to follow in a natural
way from a change in the concept of arena on which physical processes
occur, i.e. from a shift from the ordinary 3D 
 space to the 6D phase space. 
In
our scheme, the two rishons $V$ and $T$ correspond precisely to
two eigenvalues ($-1/3,+1/3$) of the ``partial hypercharge''
$Y_k=\sigma_k\otimes\sigma_k\otimes \sigma_3/3$ that emerges
 from the consideration
of phase-space transformations.
The value of $k=1,2,3$ corresponds to the position of rishon in the
Harari-Shupe model. 
Thus, for any $k$, rishons $V$ and $T$ 
are just two different eigenvalues of a {\it single} algebraic entity.
All this explains why the ordering of rishons
is important, leads to the SU(3) colour degree of freedom, 
and removes the arbitrariness present
in the original Harari-Shupe scheme. 
Since each rishon corresponds
 to just one direction in the ordinary 3D world,
 the concept of spin cannot be applied to it: individual rishons do not
 possess spin. The very idea of ``dividing'' loses its original meaning.\\

While getting rid of several drawbacks of the Harari-Shupe scheme,
 our approach
 clearly has its problems.
 In particular, we have not proposed any explicit link to the gauge
principle.
In that respect, therefore, 
we have not yet improved on the original ideas of Harari. He speculated
that the gauge structure is absent at the rishon level, but emerges
at the composite level, writing in \cite{Harari} 
that the dynamics at the rishon level ``should
somehow reproduce currently accepted theories''.
In fact, Harari suggested that gauge bosons are composed of rishons as well.
In our scheme, however, quarks and leptons are not composite objects at all,
i.e. they are definitely point-like when viewed in the standard 3D framework.
Neither our quarks nor leptons have any internal structure in the ordinary sense,
 and the same is
expected of gauge bosons.
Thus, our model is in fact not a preon model at all.
It just provides a possible explanation of 
the symmetry between quarks and leptons, 
as identified by Harari and Shupe, but without
any subparticle structure.
It shows that our tendency to explain such a symmetry in terms of ``preons''
may be misleading.
Obviously, our gauge bosons have to possess 
symmetry properties corresponding to those of the phase space. 
However, in order to
deal with the gauge bosons and be internally consistent, one needs to address the issue of
gauge invariance in the phase-space language (see, e.g.
\cite{Zachos}).
In our opinion, the problem here is related to a general difficulty in joining
different descriptions, often formulated at different levels, and possibly involving completely
different formalisms (or ``dynamics'' as Harari put it). 
We believe, however, that symmetry survives the change 
of description formalism (as e.g. rotation symmetry does in the transition from
the classical to quantum description), and therefore we think  
that the origin of the SM symmetry
group lies in  
the symmetries of phase space
(or else is intimately related to them).

Another problem is that, although 
in our approach weak isospin
is automatically connected with the lack of invariance under reflections,
there does not seem to be a strict correspondence to the pattern
of parity violation built into the Standard Model.
Then there is the problem of mass (including the question of the existence of
Higgs particle). With both the Planck constant and the
new constant of dimension [momentum/position] needed in our approach, 
the natural mass scale is set when the velocity of light $c$ is
added. It seems therefore that one should expect the
scheme to be able to say something about mass.
In fact, the issue of mass constituted one of the
questions from which our approach originally started, and
some symmetry-based conjectures have already been made
\cite{APPB2007a,APPB2007b}. 
While a more explicit proposal (presumably at the level of
phase-space-induced algebra) is still missing, 
we hope that our approach has the potential
to provide a different angle on the problem of mass.\\

The general idea behind our scheme is 
that (at least some of) the internal symmetries 
built into the Standard
Model 
and the related quantum numbers
represent an image of the
symmetries of nonrelativistic quantum phase space
(or underlie these symmetries).
This idea is in strict analogy to the well-known
connection between spin (parity) and the symmetry
properties of ordinary 3D space.
The presented proposal constitutes a kind of ``minimal solution'', 
in which a simple mathematical structure 
realizes and reflects
the basic physico-philosophical idea.
A better description of the real world is expected to require
a variation on the theme.
If the origin of internal symmetries is indeed connected with phase-space
properties, then a better understanding of our macroscopic world 
should follow
from the studies of elementary particles.

\vfill

\vfill

\end{document}